\documentclass[%
 reprint,
 amsmath,amssymb,
 aps,
]{revtex4-2}

\usepackage{appendix}
\usepackage{graphicx}
\usepackage{dcolumn}
\usepackage{bm}

\begin{document}

\preprint{APS/123-QED}

\title{
Quantum speed limit under decoherence: unitary, dissipative, and fluctuation contributions
}

\author{Kohei Kobayashi}
 \altaffiliation[Also at ]{
 Global Research Center for Quantum Information Science, National Institute of Informatics,
\\ 2-1-2, Hitotsubashi, Chiyoda-Ku,Tokyo 101-8430, Japan
 }
 \email{k\_kobayashi@nii.ac.jp}
\affiliation{Global Research Center for Quantum Information Science, National Institute of Informatics,
 2-1-2, Hitotsubashi, Chiyoda-Ku,Tokyo 101-8430, Japan}


\date{\today}

\begin{abstract}
We derive a new quantum speed limit (QSL) for open quantum systems governed by Markovian dynamics.
By analyzing the time derivative of the Bures angle between the initial pure state and its time-evolved state,
we obtain an analytically computable upper bound on the evolution speed that decomposes into three distinct physical contributions; coherent unitary dynamics, dissipative deformation, and a fluctuation term.
Based on this structure, we establish a general inequality that connects the QSL to the Quantum Fisher information in the short-time regime.
This result gives a fundamental trade-off between the distinguishability between speed and estimation precision, and clarifies how decoherence can both accelerate and constrain information acquisition.

\end{abstract}

\maketitle


\section{Introduction}

Controlling the dynamics of quantum systems with high precision is a central challenge in modern quantum science, with applications ranging from quantum computing and sensing to thermodynamics.
A fundamental constraint in this context is the quantum speed limit (QSL), which sets a theoretical bound on how quickly a quantum system can evolve from one state to another. 
While QSLs were originally formulated for isolated systems based on energy-time uncertainty relations \cite{MT, ML, Fleming, Bhattacharyya, Pfeifer, Giovanetti}, recent work has extended the concept to open quantum systems governed by dissipative dynamics \cite{Taddei, Campo, Deffner, Sun, Meng, Mirkin, Zhang, Funo, Shiraishi, Kohei}.

These generalized QSLs are not only of fundamental interest but also play a crucial role in evaluating the performance limits of practical quantum devices, 
such as quantum gates \cite{Caneva, Arenz, Lee}, sensors \cite{Herb}, and feedback-controlled systems \cite{Yunoki, Hou}. 
Geometric approaches based on the distinguishability between quantum states, 
such as the Bures angle \cite{Taddei} or relative purity \cite{Campo, Kohei}, 
have proven particularly insightful. 

However, many of the QSL results for open systems developed so far are rather formal and mathematically involved. 
They often depend on abstract quantities, such as operator norms or time-integrated measures, which are difficult to interpret physically. 
Because of this, it has been challenging to understand what actually limits the speed of quantum evolution in open systems. 
Recently, there has been increasing interest in developing QSL that not only give valid bounds, but also help us understand the physical roles \cite{Funo, Shiraishi}.

A notable contribution in this direction is the result by Funo \cite{Funo}, who derived a quantum speed limit for open quantum systems governed by Lindblad dynamics.
Their approach connects the minimal evolution time to thermodynamic quantities such as entropy production and energy fluctuations, and further identifies an additional quantum contribution beyond the classical thermodynamic bound.
This formulation emphasizes the role of shortcut-to-adiabaticity protocols and provides a thermodynamic interpretation of quantum coherence under dissipation.

In contrast, our approach provides a geometrically formulation of the quantum speed limit based on the Bures angle. 
By analyzing its differential change under Lindblad dynamics, we derive an explicit speed bound that decomposes the contributions from the coherent unitary evolution, the structural deformation caused by dissipation, and a fluctuation-like term associated with quantum coherence.

Importantly, this geometric formulation enables a direct and rigorous connection between the quantum speed limit and the quantum Fisher information (QFI). 
This result gives a fundamental trade-off between the speed of quantum evolution and the precision of parameter estimation. 
In particular, it shows that information acquisition through dissipative dynamics is intrinsically limited by how fast the state departs from its initial configuration.

\section{Main result}

\subsection{Geometric upper bound on state distinguishability under open dynamics}

We consider the open quantum system obeying the master equation:

\begin{align}
    \frac{d\rho_t}{dt}=-i[H, \rho_t]+\mathcal{D}[L]\rho_t,
\end{align}
where $H$ is the Hamiltonian and $L$ is the Lindblad operator corresponding the decoherence.
Hence, $\mathcal{D}[L]\rho=L\rho L^\dagger-L^\dagger L\rho/2-\rho L^\dagger L/2$.
We assume throughout this paper that $H$ and $L$ are time-independent, and initial state is pure
$\rho_0=|\psi_0\rangle\langle\psi_0|$ and $\hbar=1$.
Next, we define the Bures angle between $\rho_0$ and $\rho_t$ as

\begin{align}
 \Theta_t={\rm arccos}\left\{ \sqrt{{\rm Tr}(\rho_0\rho_t)}\right\}.
\end{align}
It satisfies $\Theta_t\in[0, \pi/2]$.
The maximum is achieved when $\rho_t$ is orthogonal to $\rho_0$, and 
the minimum is achieved only when $\rho_t=\rho_0$.
Here, we derive an upper bound of the time evolution of $\Theta_t$.
This derivation follows the technique introduced in our earlier work \cite{Kohei}.

First, the dynamics of $\rho_t$ is given by

\begin{align}
\label{dynamics}
&\frac{d\Theta_t}{dt}
    =\frac{-1}{\sqrt{1-{\rm Tr}(\rho_0\rho_t)^2 }}\frac{1}{2\sqrt{{\rm Tr}(\rho_0\rho_t) }}
 {\rm Tr}\left(\rho_0 \frac{d\rho_t}{dt}\right)    \notag \\
 &=\frac{1}{2\sin\Theta_t\cos\Theta_t}
 \left( {\rm Tr}\{i[\rho_0, H]\rho_t\} 
 - {\rm Tr}\{\rho_t\mathcal{D}^\dagger[L]\rho_0\} \right),
\end{align}
where $\mathcal{D}^\dagger[L]\rho=L^\dagger \rho L-L^\dagger L\rho/2-\rho L^\dagger L/2$.
To obtain a upper bound of the righthand side of Eq. (\ref{dynamics}), we use two inequalities.
One is the Cauchy-Schwarz inequality for matrices $X$ and $Y$,

\begin{align}
|{\rm Tr}(X^\dagger Y)|\le \|X\|_{\rm F}\|Y\|_{\rm F},
\end{align}
where $\|X\|_{\rm F}=\sqrt{{\rm Tr}(X^\dagger X)}$ is the Frobenius norm.
The other one is 

\begin{align}
\|\rho_t-\rho_0\|^2_{\rm F} &={\rm Tr}[(\rho_t-\rho_0)^2] \notag \\
&={\rm Tr}(\rho_t^2-2\rho_t\rho_0 + \rho_0^2) \notag \\
&\le 2-2{\rm Tr}(\rho_t\rho_0) \notag \\
&=2\sin\Theta_t,
\end{align}

where we used ${\rm Tr}(\rho_t^2)\le 1$ and ${\rm Tr}(\rho_0^2)= 1$.
Using these inequalities, the righthand side of  Eq. (\ref{dynamics}) is bounded as follows:

\begin{align}
 {\rm Tr}\{i[\rho_0, H]\rho_t\} 
&={\rm Tr}\{i[\rho_0, H] (\rho_t-\rho_0)\} \notag \\
&\le \| i[H, \rho_0]\|_{\rm F} \|\rho_t-\rho_0\|_{\rm F}  \notag \\
&\le 2 \Delta H_0\sin\Theta_t,  
\end{align}
where $\Delta H_0=\sqrt{{\rm Tr}(H^2\rho_0)-{\rm Tr}(H\rho_0)^2}$ is the energy variance of the initial state. And also
\begin{align}
& - {\rm Tr}\{\rho_t\mathcal{D}^\dagger[L]\rho_0\} \notag \\
&=- {\rm Tr}\{(\rho_t-\rho_0)\mathcal{D}^\dagger[L]\rho_0\} - {\rm Tr}\{\rho_0 \mathcal{D}^\dagger[L]\rho_0\} \notag \\
&\le \|\mathcal{D}^\dagger[L]\rho_0 \|_{\rm F} \|\rho_t-\rho_0\|_{\rm F} +
{\rm Tr}(L^\dagger L\rho_0)-{\rm Tr}(L^\dagger\rho_0 L\rho_0)  \notag \\
&\le \sqrt{2}\|\mathcal{D}^\dagger[L]\rho_0 \|_{\rm F}\sin\Theta_t + \|L|\psi_0\rangle\|^2-|\langle \psi_0|L|\psi_0\rangle|^2 \notag \\
&\le \sqrt{2}\mathcal{G}\sin\Theta_t +  \mathcal{E},
 \end{align}
where we wrote $\mathcal{G}=\|\mathcal{D}^\dagger[L]\rho_0 \|_{\rm F}$ and 
$\mathcal{E}= \|L|\psi_0\rangle\|^2-|\langle \psi_0|L|\psi_0\rangle|^2$, and
$\||\psi\rangle\|^2=\langle \psi|\psi\rangle$ is the Euclidean norm.

Combining these, we have 
\begin{align}
\label{upperbound}
    \frac{d\Theta_t}{dt}
\le \frac{1}{\sin2\Theta_t}
 \left(2\Delta H_0 \sin\Theta_t
 + \sqrt{2}\mathcal{G}\sin\Theta_t 
 +\mathcal{E} \right).
\end{align}

The inequality above provides a geometric upper bound on the rate of change of the Bures angle between the initial pure state $\rho_0$ and the time-evolved state $\rho_t$ under Markovian dynamics. 
It clearly separates the contributions of three distinct physical origins.
The first term $\Delta H_0$ represents the unitary part of the evolution and reduces to the standard Mandelstam–Tamm bound in the absence of decoherence.
The second term $\mathcal{G}$ quantifies the structural deviation induced by the dissipative channel and can be interpreted as a measure of how strongly the environment makes the state away from its original direction in state space. 
The final term $\mathcal{E}$ corresponds to the quantum variance of the Lindblad operator $L$ with respect to the initial state. 
Together, these contributions determine how fast the state departs from its initial configuration under the combined influence of coherent and incoherent dynamics. 
Importantly, all terms in the bound are expressed in terms of the initial state and known dynamical generators, and thus it becomes possible to analytically compute and physically interpret the speed limit.

\subsection{Quantum speed limit}

Importantly, the inequality (\ref{upperbound}) can be integrated analytically to yield an explicit bound on the evolution time.
By integrating this from $t=0$ to $T$, we obtain following expression for the quantum speed limit:

\begin{align}
\label{qsl}
T \ge T_{\rm QSL}
:=\frac{2}{\mathcal{V}}
\left( \sin\Theta_T -\frac{\mathcal{E}}{\mathcal{V}}
\ln\left[1+ \frac{\mathcal{V}}{\mathcal{E}} \sin\Theta_T\right]\right),
\end{align}
where we define the effective speed coefficient as
\begin{align}
\mathcal{V} =2\Delta H_0+\sqrt{2}\mathcal{G}.
\end{align}

This inequality represents the interplay between unitary dynamics, dissipative structure, and quantum fluctuations in the system's evolution.
The term $\mathcal{V}$
acts as an effective evolution rate, combining the energy variance of the initial state and a geometric contribution $\mathcal{G}$
measuring the contribution of dissipative deformation that changes the distinguishability between quantum states through nonunitary effects.
The logarithmic function involving 
$\mathcal{E}$ reflects the influence of fluctuation-like effects arising from the quantum variance of the Lindblad operator with respect to the initial state.

This form of the bound clearly reveals two distinct physical regimes.
When the decoherence is small, i.e., $L\to0$, the evolution speed is governed primarily by the effective coherent contributions, recovering a Mandelstam–Tamm–like behavior: 
\begin{align}
T_{\rm QSL}\to\frac{\sin\Theta_T}{\Delta H_0}.
\end{align}

In contrast, in the strong decoherence regime $\mathcal{E} \to \infty$,
by using the Taylor expansion $\ln(1+x)\approx x-x^2/2$ for $x\ll1$,
the bound simplifies to
\begin{align}
T_{\rm QSL} &\approx 
\frac{2}{\mathcal{V}}\left(\sin\Theta_t-\frac{\mathcal{E}}{\mathcal{V}}\left[\frac{\mathcal{V}}{\mathcal{E}}\sin\Theta_t-\frac{1}{2}\left(\frac{\mathcal{V}}{\mathcal{E}}\right)^2 
\right]\right)   \notag   \\
&\approx \frac{\sin^2\Theta_T}{\mathcal{E}}.
\end{align}
This implies that distinguishability between quantum states can be achieved in arbitrarily short time. 
This acceleration does not originate from coherent evolution, but rather from rapid structural deformation and decoherence-induced diffusion in the state space.
Unlike the conventional quantum Zeno effect, where strong coupling inhibits evolution, this result highlights the opposite phenomenon: dissipation can serve as a resource for fast state transformation.

In practical open systems, both the effective evolution rate $\mathcal{V}$ and the fluctuation-like term $\mathcal{E}$ typically scale with the dissipation strength $\gamma$. 
Assuming $\mathcal{V}/\mathcal{E} = r$ remains constant in the strong decoherence limit, the bound reduces to
\begin{align}
& T_{\mathrm{QSL}} \sim \frac{1}{\mathcal{V}} \cdot f(r, \Theta_T), \\
&f(r, \Theta_T) := 2\left( \sin\Theta_T - \frac{1}{r} \ln(1 + r \sin\Theta_T) \right).   
\end{align}

This implies that $T_{\mathrm{QSL}}$ still vanishes as $\gamma \to \infty$, but the rate is governed not by the fluctuation strength $\mathcal{E}$ alone, but by the combined scaling of $\mathcal{V}$ and $\mathcal{E}$. 
The ratio $r$ represents the relative contribution of structural deformation to dissipative fluctuations. 
Therefore, dissipation can act both as a driver of distinguishability and as a constraint on the speed limit, depending on the structure of the Lindblad operator.

\section{Example}

\subsection{Qubit}
To illustrate the features of the derived QSL, we consider a simple one-qubit system consisting of the excited state $|0\rangle=(1, 0)^\top$ and the ground state $|1\rangle=(0, 1)^\top$, subject to unitary evolution and pure dephasing.
We take the system's initial state to be a pure qubit state
\begin{align}
\label{initial}
|\psi_0\rangle=\cos\frac{\theta}{2}|0\rangle +\sin\frac{\theta}{2}|1\rangle,
\end{align}
where $\theta\in[0, \pi]$ characterizing the superposition. 
The Hamiltonian is chosen as
\begin{align}
H=\frac{\omega}{2}\sigma_x
=\frac{\omega}{2}(|0\rangle\langle1|+|1\rangle\langle0|),
\end{align}
which represents the coherent Rabi oscillations about the $x$-axis.
Decoherence of dephasing is modeled by
\begin{align}
L=\sqrt{\gamma}\sigma_z=\sqrt{\gamma}(|0\rangle\langle0|-|1\rangle\langle1|).
\end{align}

In this case, the quantities in our QSL can be evaluated analytically:
\begin{align*}
\Delta H_0 &=\frac{\omega}{2}\sqrt{{\rm Tr}[\sigma^2_x\rho_0]-{\rm Tr}[\sigma_x\rho_0]^2 }  \\
&=\frac{\omega}{2} \cos\theta, \\
\|\mathcal{D}^\dagger[L]\rho_0\|_{\rm F} 
&=\gamma\|\sigma_z\rho_0\sigma_z -\frac{1}{2}\sigma_z^2\rho_0
-\frac{1}{2}\rho_0\sigma_z^2\|_{\rm F}  \\
&= 2\gamma \sin\theta, \\
\mathcal{E} &=\gamma\|\sigma_z|\psi_0\rangle\|-\gamma|\langle \psi_0|\sigma_z|\psi_0\rangle|^2     \\
&= \gamma \sin^2\theta,
\end{align*}
and the combined rate coefficient is given by
\begin{align}
\mathcal{V} &= 2\Delta H_0 + \sqrt{2}\mathcal{V} \notag \\
&= \omega \cos\theta + 2\sqrt{2}\gamma \sin\theta.
\end{align}

By substituting these into the QSL yields
\begin{equation}
\label{eq:qubit_qsl}
T_{\mathrm{QSL}} = \frac{2}{\mathcal{V}} \left( \sin\Theta_T - \frac{\mathcal{E}}{\mathcal{V}} \ln\left[1 + \frac{\mathcal{V} \sin\Theta_T}{\mathcal{E}} \right] \right).
\end{equation}

We fix $(\theta, \Theta_T) = (\pi/4, \pi/4)$ and evaluate $T_{\mathrm{QSL}}$ as a function of the dissipation strength $\gamma$ for various values of $\omega$.

Figure 1(a) shows the behavior of the QSL. 
In the weak dissipation limit $\gamma \to 0$, the QSL is dominated by the coherent contribution $\Delta H_0$, and $T_{\mathrm{QSL}}$ decreases with increasing $\omega$, consistent with Mandelstam–Tamm-like behavior. 
As $\gamma$ increases, the contributions from $\|\mathcal{D}^\dagger[L]\rho_0\|_{\rm F}$ and $\mathcal{E}$ become significant, 
accelerating the state evolution.

In the strong dissipation regime, both $\mathcal{V}$ and $\mathcal{E}$ scale linearly with $\gamma$; then we find $T_{\mathrm{QSL}} \propto 1/\gamma$. 
This indicates that decoherence can make the state evolution faster, not by rotating the state coherently but by deforming it irreversibly.
This behavior contrasts with the conventional quantum Zeno effect and highlights the constructive role of dissipation in open dynamics.

Next, as an important topic, we examine 
the tightness of the derived bound.
We consider a simple yet analytically tractable model;
the qubit is subjected to the spontaneous emission modeled by 
\(L = \sqrt{\gamma}\sigma_-=\sqrt{\gamma}|1\rangle\langle0|\), with no Hamiltonian \(H = 0\).  
We choose the initial state to be the excited state $|\psi_0\rangle = |0\rangle$.
Under this dynamics, the quantum state evolves as
\[
\rho_t= 
\begin{pmatrix}
e^{-2\gamma t} & 0 \\
0 & 1 - e^{-2\gamma t}
\end{pmatrix},
\]
and the fidelity with respect to the initial state is $F_t=\langle\psi_0|\rho_t|\psi_0\rangle=  e^{-2\gamma t}$.  
Hence, the Bures angle is given by
\[
\Theta_t = \arccos(e^{-\gamma t}).
\]
Solving for the time \(T_{\mathrm{exa}}\) required to reach a given angle \(\Theta_T\),
we obtain the exact expression:
\[
T_{\mathrm{exa}} = \frac{-1}{\gamma} \ln\left[\cos ^2\Theta_T\right].
\]

On the other hand,  with $\mathcal{V}= \sqrt{2}\gamma$ and $\mathcal{E} = \gamma$.
the QSL yields
\[
T_{\mathrm{QSL}} = \frac{\sqrt{2}}{\gamma} \left( \sin\Theta_T - \frac{1}{\sqrt{2}} \ln\left(1 + \sqrt{2} \sin\Theta_T \right) \right).
\]

Figure 1(b) compares \(T_{\mathrm{exa}}\) and \(T_{\mathrm{QSL}}\) as functions of the target angle \(\Theta_T\).
The results show that \(T_{\mathrm{QSL}}\) is 
always less than or equal to $T_{\mathrm{exa}}$, confirming the validity of the derived bound.
When $\Theta_T$ is sufficiently small, the two plots converge, indicating that the bound is asymptotically tight.
As \(\Theta_T\) increases and approaches $\pi/2$, 
\(T_{\mathrm{exa}}\) diverges due to the vanishing overlap between initial and final states, while \(T_{\mathrm{QSL}}\) remains finite.
This behavior means the fact that amplitude damping cannot fully transfer population from the excited state to the ground state in finite time.
These results demonstrate that our bound not only respects the fundamental constraints of open quantum dynamics but also provides a tight and computable estimate of the minimal evolution time.

Notably, in the amplitude damping model considered here, both the exact evolution time and the QSL scale inversely with the dissipation rate 
$\gamma$. 
As a result, their ratio $T_{\rm exa}/T_{\rm QSL}$ is independent of $\gamma$ and depends solely on the target Bures angle.
This reflects the fact that the evolution speed is entirely governed by the dissipative rate in the absence of coherent dynamics, making the QSL an effective and physically interpretable bound for speed of system's dynamics.

\begin{figure}[tb]
\centering
\includegraphics[width=0.9\linewidth]{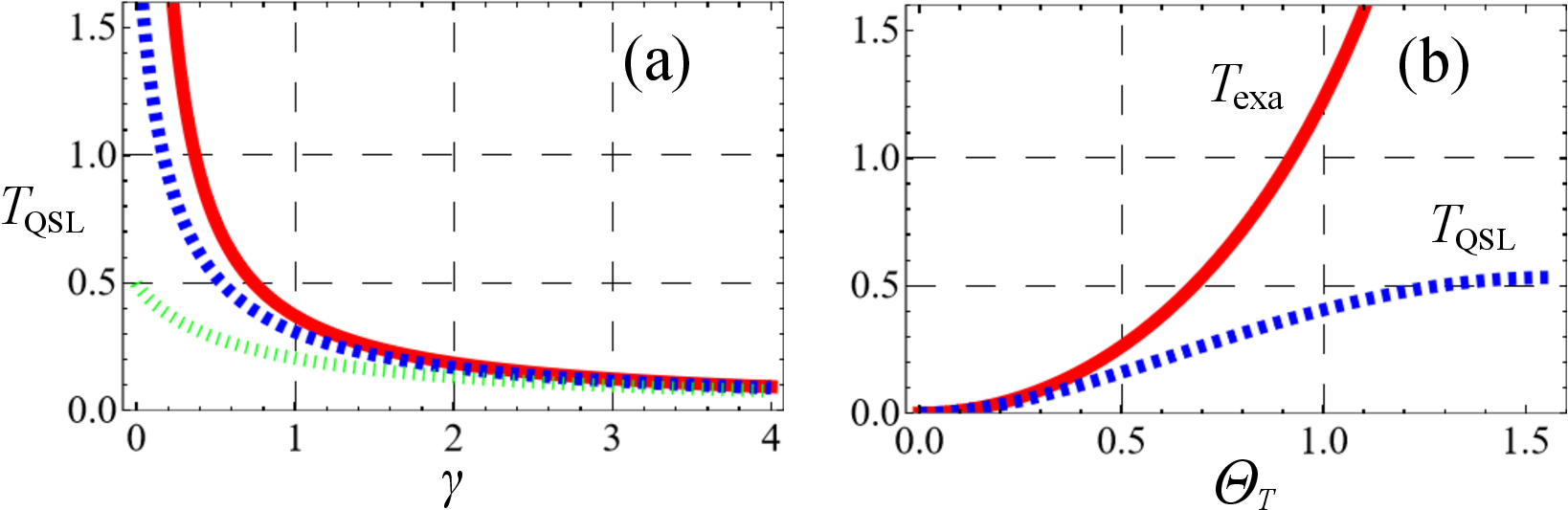}
\caption{(a) $T_{\rm QSL}$ as a function of $\gamma$ for $\omega=0.01$ (red solid line), 
$\omega=1$ (blue dashed line), and $\omega=4$ (green dotted line).
The parameters are fixed $(\theta, \Theta_T)=(\pi/4, \pi/4)$.
(b) Comparison of $T_{\rm exa}$ and $T_{\rm QSL}$ as a function of $\Theta_T$, with fixed
$\gamma=1$ and $\theta=\pi/4$. }
\end{figure}

\subsection{Many-Body system}

We consider an $N$-qubit system, whose dynamics is generated by
\begin{align}
H=\sum_{i=1}^{N}\frac{\omega}{2}\sigma_z^{(i)}, 
\end{align}
where $\sigma_z^{(i)}=I^{(1)}\otimes \cdots \otimes \sigma^{(i)}_z \otimes \cdots \otimes I^{(N)}$, and
\begin{align}
L=\sqrt{\gamma}\sigma_x^{(i)}.
\end{align}

$H$ acts on the $i$th qubit respectively and $L$ acts on the $i$th qubit only.
Due to the local nature of both the Hamiltonian and the dissipation, the contributions to the energy variance, geometric deformation, and fluctuation-like terms scale with system size in a tractable way. 
This allows for a clear scaling analysis of the QSL in the large-$N$ limit.

The initial state is taken to be a product state of the pure state $\rho_0=|\psi_0\rangle \langle \psi_0|$:
\begin{align}
\rho_0 = \rho_0^{(1)} \otimes \rho_0^{(2)}\otimes \cdots \otimes \rho_0^{(N)}.
\end{align}
with single-qubit Bloch polar angle $\theta$.
Although this model is still idealized, it shows the competition
between coherent rotation and local dephasing.
Since the variance of $H$ in each spins are independent, we obtain
\begin{align}
\Delta H^2_0 
= \left(\frac{\omega}{2}\right)^2\sum^N_{i=1}\Delta \sigma^{(i)}_{z, 0}
=\mathcal{O}(N),
\end{align}
and therefore
\begin{align}
2\Delta H_0=\mathcal{O}(\sqrt{N}).
\end{align}
The terms $\mathcal{G}$ and 
$\mathcal{E}$ are given by
\begin{align}
\mathcal{G} &=2N \gamma\sin\theta = \mathcal{O}(N\gamma),   \notag \\
\mathcal{E} &= N\gamma\sin^{2}\theta =\mathcal{O}(N\gamma).
\end{align}

Hence, in the dissipation-dominated regime ($\gamma\gg\omega$),
the ratio
$r=\mathcal{V}/\mathcal{E}$ approaches the constant  
\(r\simeq 2/\sin\theta\).
Substituting these scalings into the bound yields
\begin{align}
\label{N_scaling}
T_{\mathrm{QSL}}
&\sim \frac{2}{N\gamma}
\left(\sin\Theta_T-\frac{1}{r}\ln\bigl[1+r\sin\Theta_T\bigr]
\right)  \notag \\
&\propto\frac{1}{N\gamma}.
\end{align}

Eq. (\ref{N_scaling}) indicates a $1/N$ speed-up.
This means that adding more qubits makes the system evolve faster, as long as each qubit experiences the same local dissipation.
This speed-up does not come from quantum entanglement, but from the fact that each qubit independently interacts with its own environment.
In this way, dissipation acts like an information flow that drives the system away from its initial state, making the evolution faster as the system becomes larger.
This scaling behavior remains valid even if weak local interactions are added, as long as they do not create strong correlations between qubits.
On the other hand, if the dissipation is global and acts on the whole system collectively, the speed limit no longer scales with the number of qubits.
This shows that the structure of the Lindblad operators plays a key role in determining how fast the system can evolve.

\section{Quantum speed limit and Fisher information}

The quantum Fisher information (QFI) $F_Q$ characterizes the ultimate precision limit for estimating a parameter encoded in a quantum state. In the context of time evolution, it quantifies how rapidly a state becomes distinguishable from its initial one.

We begin by recalling the relationship between the Bures angle and the quantum Fisher information in the short-time regime.
It is well known that the quantum Fisher information is connected with the Fidelity as follows \cite{Taddei, Braunstein}:
\begin{align}
F(\rho_0, \rho_T) = 1 - \frac{F_Q}{4} T^2 + o(T^3).
\end{align}
Using it and the Taylor expansion of 
$\arccos(\sqrt{1 - \epsilon}) \approx \sqrt{\epsilon}$ for small $\epsilon > 0$, we obtain
\begin{align}
\Theta_T &\approx \arccos\left(\sqrt{1 -F(\rho_0, \rho_T)}\right) \notag \\
 &\approx \frac{\sqrt{F_Q}}{2} T.
\end{align}
Thus, for sufficiently short times, the Bures angle grows linearly with time and the coefficient is determined by the square root of the quantum Fisher information.
Next, we consider the following inequality:
\begin{align}
\ln(1 + x) \le \frac{x(x+2)}{2(1 + x)},
\end{align}
for all $x>0$.
By applying this to the QSL expression (\ref{qsl}),
we obtain the following lower bound:
\begin{align}
T_{\mathrm{QSL}} \ge \frac{\sin^2 \Theta_T}{\mathcal{E} + \mathcal{V} \sin \Theta_T},
\label{QSL-FQ}
\end{align}
which is valid for all $\Theta_T \in (0, \pi/2)$ and $\mathcal{E}, \mathcal{V}$. 

Substituting the short-time relation 
$\sin \Theta_T \approx \Theta_T\approx\sqrt{F_Q} T/2$ into Eq. (\ref{QSL-FQ}),
we have
\begin{align}
T \ge \frac{ \frac{F_Q}{4} T^2 }{ \mathcal{E} + \frac{\mathcal{V}}{2} \sqrt{F_Q} T }.
\end{align}
Rearranging terms yields a quadratic inequality in $\sqrt{F_Q}$, which leads to the following upper bound on the Fisher information:
\begin{align}
\label{FQ-bound}
F_Q \le \left( \mathcal{V} + \sqrt{ \mathcal{V}^2 + \frac{4 \mathcal{E}}{T} } \right)^2.
\end{align}

The upper bound gives a fundamental trade-off between the speed of quantum evolution and the attainable estimation precision. 
In particular, for sufficiently short times, the bound diverges as $F_Q\lesssim \mathcal{O}(1/T^2)$, reflecting the fact that information cannot be extracted arbitrarily quickly due to the limited distinguishability between quantum states. 
Furthermore, since the upper bound depends explicitly on the effective speed $\mathcal{V}$ and the dissipative fluctuation term $\mathcal{E}$, it clarifies how coherent dynamics and decoherence jointly constrain the precision of parameter estimation. 
This result provides an operationally meaningful generalization of the quantum Cramér–Rao bound to dissipative quantum systems and indicates that dissipation, while often seen as obstacle, can also act as a source of information acquisition under appropriate conditions.

\section{Conclusion}

We have derived a speed limit $T_{\rm QSL}$ for Markovian open quantum systems obeying Lindblad-type dynamics, where the evolution time is explicitly bounded in terms of the Bures angle between the initial and final states. 
Our formulation decomposes the generator of state change into three physically interpretable contributions: a unitary term governed by the energy variance, a geometric term reflecting the structural deformation induced by dissipation, and a fluctuation term that quantifies the quantum variance of the Lindblad operator with respect to the initial state.

We have demonstrated the utility of our bound using analytically tractable qubit models and simple $N$-spin systems with local dephasing. 
In these examples, the QSL provides a tight upper bound on the achievable distinguishability, and reveals that dissipation can accelerate state evolution in proportion to system size not by coherent entanglement, but by enhancing the rate of information flow out of the system.

Furthermore, we established a direct connection between the QSL and the quantum Fisher information, showing that the latter is upper bounded by a function of the effective speed 
coefficient $\mathcal{V}$ and $\mathcal{E}$. 
This relation gives an intuition on how the precision of quantum parameter estimation is fundamentally constrained by the structure of the generator of dynamics. 
This bound holds in the short-time approximation and provides a robust operational limit that is especially relevant in noisy quantum systems. 
Our result demonstrates that the speed at which a quantum state departs from its initial state intrinsically limits the amount of information that can be extracted from it in finite time.

Our result provides new insights into the fundamental limitations and capabilities of quantum control in the presence of noise, from both geometric and information-theoretic perspectives.
Future works is to extend this approach to more general noise models, time-dependent controls, or estimation-based feedback protocols, further deepening the connection between quantum speed, distinguishability, and information acquisition in open quantum systems.

This work was supported by MEXT Quantum Leap Flagship Program Grant JPMXS0120351339.



\begin{thebibliography}{99}

\bibitem{MT}
L. Mandelstam and I. Tamm, The Uncertainty Relation Between Energy and Time in Non-relativistic Quantum Mechanics, Izv. Akad. Nauk SSSR {\bf 9}, 122 (1945)
[J. Phys. (USSR) {\bf 9}, 249 (1945)].

\bibitem{ML}
N. Margolus and L. B. Levitin, The maximum speed of dynamical evolution,
Physica D {\bf 120}, 188 (1998).

\bibitem{Fleming}
G. N. Fleming, 
A unitarity bound on the evolution of nonstationary states,
Nuovo Cimento A (1965-1970) {\bf 16}, 232 (1973).

\bibitem{Bhattacharyya}
K. Bhattacharyya,
Quantum decay and the Mandelstam-Tamm-energy inequality,
J. Phys. A: Math. Gen. {\bf 16}, 2993 (1983).


\bibitem{Pfeifer}
P. Pfeifer,
How fast can a quantum state change with time?,
Phys. Rev. Lett. {\bf 71}, 306 (1993).

\bibitem{Giovanetti}
V. Giovannetti, S. Lloyd, and L. Maccone,
The quantum speed limit,
Phys. Rev. A {\bf 67}, 052109 (2003).



\bibitem{Taddei}
M. M. Taddei, B. M. Escher, L. Davidovich, and R. L. de Matos Filho,
Quantum speed limit for physical processes,
Phys. Rev. Lett. {\bf 110}, 050402 (2013).

\bibitem{Campo}
A. del Campo, I. L. Egusquiza, M. B. Plenio, and S. F. Huelga,
Quantum speed limits in open system dynamics,
Phys. Rev. Lett. {\bf 110}, 050403 (2013).

\bibitem{Deffner}
S. Deffner and E. Lutz,
Quantum speed limit for non-Markovian dynamics,
Phys. Rev. Lett. {\bf 111}, 010402 (2013).


\bibitem{Sun}
Z. Sun, J. Liu, J. Ma, and X. Wang, 
Quantum speed limits in open systems: Non-Markovian dynamics without rotating-wave approximation,
Sci. Rep. {\bf 5}, 8444 (2015).

\bibitem{Meng}
 X. Meng, C. Wu, and H. Guo,
 Minimal evolution time and quantum speed limit of non-Markovian open systems, 
 Sci. Rep. {\bf 5}, 16357 (2015).

\bibitem{Mirkin}
N. Mirkin, F. Toscano, and D. A. Wisniacki,
Quantum Speed Limit Bounds in an Open Quantum Evolution,
Phys. Rev. A {\bf 94}, 052125 (2016).


\bibitem{Zhang}
Y.-J. Zhang, W. Han, Y.-J. Xia, J.-P. Cao, and H. Fan,
Quantum speed limit for arbitrary initial states,
 Sci. Rep. {\bf 4}, 4890 (2014).

\bibitem{Funo}
K. Funo, M. Ueda, and K. Maruyama, 
Speed limit for open quantum systems,
New J. Phys. {\bf 21}, 013006  (2019).

\bibitem{Kohei}
K. Kobayashi and N. Yamamoto,
Quantum speed limit for robust state characterization and engineering,
Phys. Rev. A {\bf 102}, 042606 (2020). 

\bibitem{Shiraishi}
N. Shiraishi and K. Saito,
Speed limit for open systems coupled to general environments,
Phys. Rev. Research {\bf 3}, 023074 (2021).


\bibitem{Caneva}
T. Caneva, M. Murphy, T. Calarco, R. Fazio, S. Montangero, V. Giovannetti, and G. E. Santoro,
Optimal Control at the Quantum Speed Limit,
Phys. Rev. Lett. {\bf 103}, 240501 (2009).


\bibitem{Arenz}
C. Arenz, B. Russell, D. Burgarth, and H. Rabitz,
The roles of drift and control field constraints upon quantum control speed limits,
New J. Phys. {\bf 19}, 103015 (2017).

\bibitem{Lee}
J. Lee, C. Arenz, H. Rabitz, and B. Russell,
Dependence of the quantum speed limit on system size and control complexity,
New J. Phys. {\bf 20}, 063002 (2018).

\bibitem{Herb}
K. Herb and C. L. Degen,
Quantum speed limit in quantum sensing,
Phys. Rev. Lett. {\bf 133}, 210802 (2024).


\bibitem{Yunoki}
H. Yunoki and  Y. Hasegawa,
Quantum Speed Limit and Quantum Thermodynamic Uncertainty Relation under Feedback Control,
arXiv:2502.09081 [quant-ph].


\bibitem{Hou}
L. Hou, B. Shao, and C. Wang, 
Quantum Speed Limit Under the Influence of Measurement-based Feedback Control,
Int. J. Theor. Phys. {\bf 62}, 47 (2023). 

\bibitem{Braunstein}
S. L. Braunstein and C. M. Caves, 
Statistical distance and the geometry of quantum states,
Phys. Rev. Lett. {\bf 72}, 3439 (1994).




\end{thebibliography}
\end{document}